**Solid solution and precipitation strengthening effects in basal slip, extension twinning and pyramidal slip in Mg-Zn alloys**

N. Li[a, b], C. Wang[c], M.A. Monclús[a], L. Yang[d*], J. M. Molina-Aldareguia[a*]

[a] IMDEA Materials Institute, C/Eric Kandel 2, 28906, Getafe, Madrid, Spain
[b] LNM, Institute of Mechanics, Chinese Academy of Sciences, Beijing 100190, China.
[c] State Key Laboratory of Solidification Processing, Northwestern Polytechnical University, 710072, Xi'an, PR China
[d] Hypervelocity Aerodynamics Institute, China Aerodynamics Research and Development Center, Mianyang 621000, China
[*] Corresponding authors. E-mails: jon.molina@imdea.org; lingwei.yang@cardc.cn

## Abstract

A high-throughput methodology is proposed, based on the combination of diffusion couples and advanced nanomechanical testing methods, to directly measure alloying effects on the critical resolved shear stress (CRSS) of individual deformation modes in Mg alloys. The methodology is tested in Mg-Zn alloys by assessing the alloying effects, up to Zn contents of 2 at.%, on basal slip, extension twining and pyramidal slip in two metallurgical condition: as-quenched, for which the Zn solute atoms remain homogenously dispersed in solid solution; and peak-aged, for which the Zn atoms form rod-shape $\beta_1'$ ($MgZn_2$) precipitates. A combined approach including micromechanical testing, transmission Kikuchi diffraction, and high-resolution transmission electron microscopy was performed to reveal the corresponding deformation mechanisms. It was found that the CRSS enhancement for basal slip and extension twinning by $Mg_2Zn$ precipitates is considerably larger than the effect of Zn in solid solution, while the strengthening of pyramidal slip is similar in both cases. As a result, the anisotropy ratios remain high and similar to pure Mg in the solid solution strengthened Mg-Zn alloys. However, they are substantially reduced in precipitation strengthened Mg-Zn alloys.

## 1. Introduction

Magnesium (Mg) has attracted significant attention among structural metallic alloys, mainly due to its lightening potential for transport and its biocompatibility for biomedical applications. However, Mg alloys maintain some drawbacks, including low strength and ductility and poor formability, as a result of their hexagonal closed packed (HCP) structure and corresponding mechanical anisotropy [1–3]. Because of this, various deformation modes, including basal, prismatic and pyramidal I slip of <a> dislocations, pyramidal II slip of <c+a> dislocations and several twinning modes are possible in Mg. However, basal slip and extension twinning are the most frequently observed because their critical resolved shear stress (CRSS), 0.5 MPa and 12 MPa, respectively, are three times lower than those for the hard prismatic and pyramidal slip systems. Therefore, overcoming the plastic anisotropy of Mg requires the use of strategies to strengthen the soft modes, such as solid solution strengthening, precipitation strengthening or grain size refinement.

In terms of the solute effects in basal slip, first principle simulations in Mg-Zn solid solutions with 0.3~0.5 at.% Zn suggested that Zn alloying increases the CRSS of basal slip more than that of non-basal slip, eventually reducing the difference in CRSS between the soft and hard slip systems [4]. For extension twinning, the solid solution effect on twinning is still the subject of controversy [5]. Stanford and Barnett [6] studied solute strengthening effects for extension twinning in binary Mg-Zn polycrystals, and concluded that the CRSS for twinning was independent of the Zn concentration up to 1 at.% (2.8 wt.%). On the contrary, Ghazisaeidi et.al [7] predicted a strengthening of around 10 MPa for AZ31 at room temperature, by computing the interaction energies of Al and Zn with the twin boundaries and the twin dislocations. Moreover, the twin nucleation may be altered by alloying atoms via the change either in the twin-boundary energy or in the atomic shuffling mechanisms [8]. Altogether, the strengthening due to solid solution effects remains relatively limited and more effective approaches for strengthening the soft modes need to be explored.

Precipitation strengthening might constitute an alternative depending on the shape, size, orientation and volume fraction of the precipitates [9–11]. The mechanisms for precipitation strengthening can be classified depending on whether the precipitates are shear-resistant or are sheared by the dislocations [12,13]. In shear-resistant particles, the increase in the CRSS for dislocation slip is attributed to Orowan mechanisms with dislocations bowing by the precipitates.

For shearable precipitates, the degree of interface coherency between the particle and the matrix plays a major role on the likelihood of a dislocation shearing the particle [13,14]. A variety of dislocation and particle interaction mechanisms might be responsible for the resistance to shearing, such as interfacial or chemical strengthening, coherency strengthening, stacking-fault strengthening, modulus strengthening and order strengthening. In Mg-Zn alloys, rod-shape $\beta_1'$ ($MgZn_2$) precipitates and plate-shape $\beta_2'$ ($MgZn_2$) are the commonly studied precipitates. Robson et.al. [9] predicted that precipitates formed on prismatic planes are more effective on strengthening Mg alloys than those formed on basal planes. For instance, rod-shape $\beta_1'$ ($MgZn_2$) precipitates, formed on the prismatic plane of Mg-Zn alloys, have shown a more effective strengthening effect than plate-shape $\beta_2'$ ($MgZn_2$), which tend to distribute in the basal planes. Moreover, the widely accepted understanding suggests that precipitates do not prevent twin nucleation, but that they might strongly interfere with the process of twin growth [13,15–19]. For instance, Robson *et al.* [20] observed that a higher density of small twins operate in age-hardened Mg-Zn alloys, than in precipitate-free samples. There is not a widely accepted view on the basic nature of the interaction between twin boundaries and precipitates [12]. Due to the complex precipitate types, the contribution of different metastable precipitates is difficult to quantify. Moreover, the strengthening provided for each type of precipitate is not known either, because most of the studies have been carried out in polycrystalline Mg-Zn alloys and it is very difficult to isolate the precipitate contribution from those coming from grain boundaries and texture effects.

In summary, there is still a limited knowledge about alloying effects on the individual deformation modes in Mg. Partly, this is because traditional approaches rely on processing a large number of alloys and testing them mechanically using conventional methods, which is costly and time-consuming. In addition, it is often difficult to extract the strengthening effect of alloying elements on individual deformation modes using these conventional methods, because the mechanical behavior is strongly influenced by other factors, like texture or grain size. Micromechanical testing is amendable to overcome these limitations, but it might also present its drawbacks. For instance, Wang and Stanford [21] quantified the precipitation effect in Mg-5.1 wt.% Zn alloys using compression of 2 μm sized pillars but did not consider size effects. Alizadeh et.al [22] studied temperature and size effects on precipitation strengthening, but they lacked a good reference because comparisons were carried out with different samples and did not have access to a continuous variation of Zn content within the same specimen.

In this work, a novel high-throughput methodology is proposed, based on the combination of diffusion couples and advanced micromechanical testing methods, to directly measure alloying effects on the CRSS of individual deformation modes in Mg alloys. This method does not only provide means of fast screening alloying effects on the different deformation modes, but also allows for direct comparison between solid-solution and precipitation hardening effects within the same grain, by subjecting the diffusion couple to different thermal treatments. In particular, in this case, the effect of Zn content on basal slip, extension twinning, and pyramidal slip was quantified within the same grain against solid solution and precipitation strengthening effects, by subjecting a Mg/Mg-Zn diffusion couple to two different thermal treatments: as-quenched (AQ) treatment, providing a supersaturated solid solution, and a peak-aged (PA) treatment, for which the formation of rod-shape $\beta_1'$ ($MgZn_2$) precipitates prevails.

## 2. Experimental methods

### *2.1 Materials and microstructural characterization*

The starting material was a Mg/Mg-2.2 at.% Zn diffusion couple, as described in previous works [23]. After diffusion annealing (400 °C for 336 h in vacuum), the diffusion couple was subjected to different thermal treatments: as-quenched (AQ) condition (350 °C for 240 h) and peak aged (PA) condition (200 ºC for 10 h), to tailor the atomic distribution of Zn in the diffusion couple. The composition line profile along the diffusion couple was measured by electron probe microanalysis (EPMA) using a JEOL Superprobe JXA 8900 microscope. The microstructure of the diffusion couple was analyzed in a dual-beam field emission gun scanning electron microscope (FEGSEM) (Helios Nanolab 600i FEI), using electron backscatter diffraction (EBSD) to determine the grain orientation and the grain size. The distribution of the Zn solute atoms in each metallurgical condition was characterized by transmission electron microscopy (TEM) using a FEI Talos F200X microscope.

### *2.2 Micropillar compression*

Micropillars with square cross-sections were machined by focused ion beam (FIB, FEI Helios Nano lab 600i) milling using the methodology reported in previous works [23]. The grain orientation was characterized by EBSD. Single crystalline square micropillars were prepared with a side-length of 5 µm to 7 µm, in order to assess potential size effects. All the pillars were prepared in the selected grain orientation using focused $Ga^+$ ion beam (FIB) milling in the same instrument (Helios NanoLab 600i, FEI). Special care was taken to ensure that the micropillars were milled at regions located at a sufficient distance away from the grain boundaries. A range of $Ga^+$ ion beam currents were employed at different milling stages, using a beam current of 40 pA for the finishing step, to minimize the surface damage due to $Ga^+$ ion implantation. Additionally, the sample was tilted $\pm$ 1.5 ° with respect to the ion beam axis during the final milling to reduce the taper. An aspect ratio between 2:1 and 3:1 (height to side length) was used in all cases with the aim of avoiding buckling effects during compression. The as-milled micropillars were carefully characterized by SEM before and after testing to determine their initial and final shape after the compression tests. The micropillars oriented for basal slip were compressed at 100 ºC to eliminate size effects using a TI950 Triboindenter (Hysitron, INC, Minneapolis, MN). For the 100 °C tests, a hot stage (Hysitron xSol) module and a diamond flat punch fitted to a special long insulating shaft were employed. The measured compressive load-displacement curves were firstly corrected by applying the Sneddon correction in order to account for the extra compliance resulting from the elastic deflection of the bulk material surrounding the micropillars. The corrected curves were then used to calculate the engineering stress (σ)-strain (ε) curves, considering the initial cross section and the height of the micropillars. From these curves, the yield stress was measured as the 0.5% offset yield stress. The Young's modulus was estimated from the upper portion of the unloading segment of the curves. The CRSS for basal slip, twin nucleation/growth and pyramidal slip were finally obtained as a function of Zn content for each metallurgical condition using the SF computed from the crystallographic orientation in each case. Fig. 1 compares the resolved shear stress-strain curves for micropillar dimensions of 5 μm ×5 μm and 7 μm×7 μm, showing a minimum size effect if the side length was larger than 5 μm. Therefore, in the following sections, only the micropillar results for a side length of 7 μm are presented. For identification purposes, the Mg-2at.%Zn in as-quenched and peak-aged condition as referred to as Mg-2Zn-AQ and Mg-2Zn-PA, respectively.

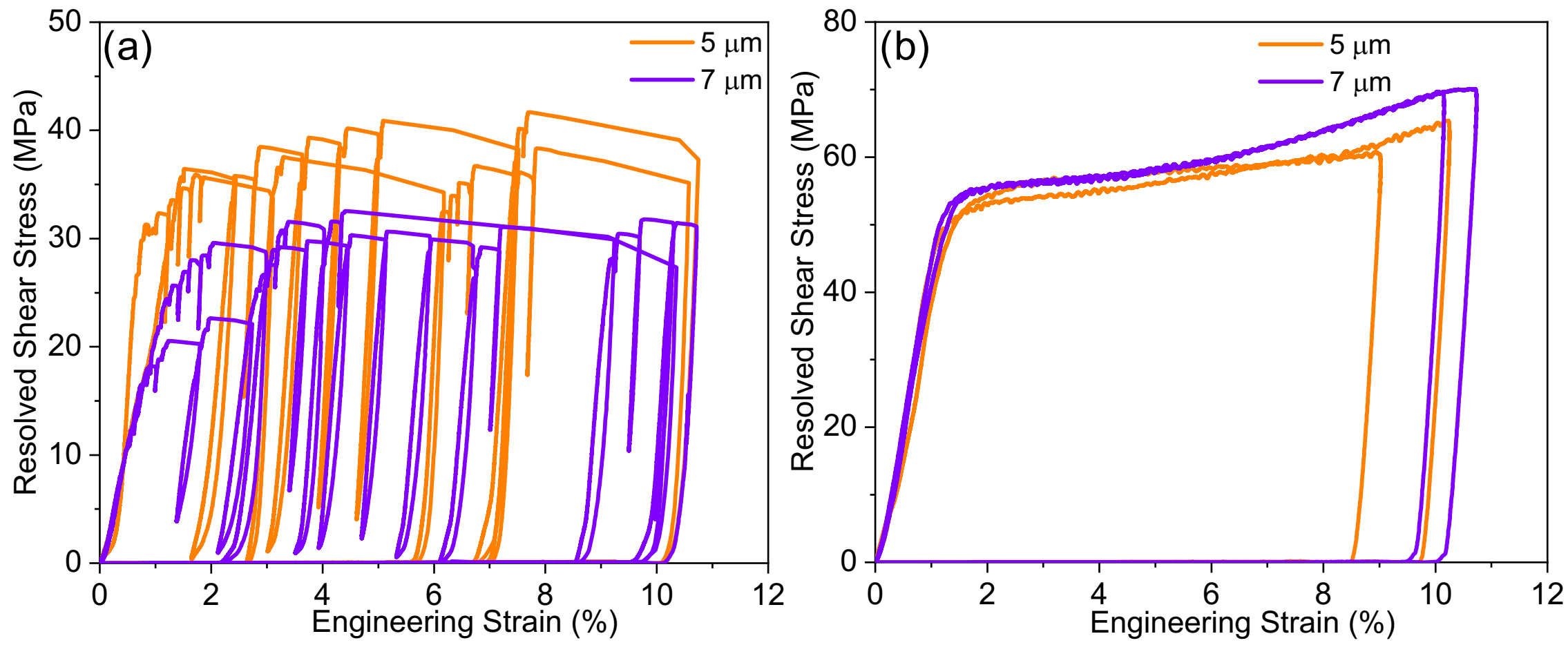

Fig. 1. Representative engineering stress- strain curves for (a) Mg-2Zn-AQ and (b) Mg-2Zn-PA condition.

Slip traces at the surface of the micropillars were carefully examined by high resolution SEM in order to determine the active slip systems. In addition, the trenching-lift FIB technique was adopted to extract lamellae, which were thinned to approximately 100 nm for electron transparency. Transmission Kikuchi Diffraction (TKD) was used to study twin activation in the same dual beam system equipped with an Oxford EBSD camera. The dislocation structures were studied by TEM. Two-beam bright-field (TBBF), two-beam dark-field (TBDF) and/or weak-beam dark-field (WBDF) imaging techniques were used to obtain dislocation contrast [24]. In addition, the 'g·b=0' invisibility criterion was used to discern the different types of dislocations present in the HCP lattice: <a> dislocations, with b $=\frac{1}{3}<11\bar{2}\,0>$,<c> dislocations with $b=[0001]$, and $<c+a>$ dislocations, with $b=\frac{1}{3}<11\bar{2}\,3>$. In particular, the $g=(0002)$ diffraction condition was employed to observe the $<c+a>$ dislocation, as $<a>$ dislocations are invisible (with 'g·b=0'), while only dislocations having the $<$ c $>$ component (i.e., <c> or <c + a> type) are in contrast.

## 3. Results and Discussion

### *3.1 Microstructure of the Mg/Mg-Zn diffusion couple*

The microstructure of Mg/Mg-Zn diffusion couple in the solid solution and peak aged conditions is shown in Fig. 2. The diffusion couple is composed of pure Mg, Mg-2at.%Zn (Mg-2Zn) and the diffusion zone in the middle with gradients of Zn ranging from 0 to 2 at.%, as shown in Fig. 2a. The grain size in the diffusion couple was of the order of 500 μm, allowing the fabrication of single-crystal micropillars in specific grains, based on the EBSD mapping shown in Fig. 2b. Grains oriented for basal slip, extension twinning, and pyramidal slip were selected for pillar preparation at both the pure Mg and Mg-2 at.% Zn sides. Table. 1 lists the maximum Schmid factors (SFs) of the different slip and twinning systems in the grains under study.

Table 1. The SFs of the grain favourably oriented for basal slip ,extension twinning and pyramidal slip in the pure Mg /Mg-2at. %Zn diffusion couple.

| Alloy | Euler angles (º) | Maximum SF | | | |
|---|---|---|---|---|---|
| | | Basal slip | Prismatic slip | Tension twinning | Pyramidal slip |
| Pure Mg | Grain 1 (87.2 ,140.5 ,28.9) | 0.43 | 0.26 | 0.17 | 0.32 |
| | Grain 2 (143.6, 81.3, 26.8) | 0.12 | 0.48 | 0.44 | 0.39 |
| | Grain 3 (167.5, 3.8, 12.6) | 0.06 | 0.49 | - | 0.47 |
| Mg-2at.%Zn | Grain 4 (30.8, 44.8 , 50.3) | 0.49 | 0.23 | 0.24 | 0.21 |
| | Grain 5 (166.9, 97.0, 21.3) | 0.11 | 0.48 | 0.48 | 0.42 |
| | Grain 6 (143.6, 81.3, 26.8) | 0.11 | 0.49 | - | 0.48 |

In the as-quenched condition, annealing at 350 °C for 240 h followed by quenching ensured a homogenous distribution of the Zn atoms without the formation of second-phase precipitates, as shown in Fig. 2c. The peak-aged heat treatment (200 ºC for 10 h) favored primarily the formation of rod-shape $\beta_1'$ ($MgZn_2$) precipitates and some secondary plate-shape $\beta_2'$ ($MgZn_2$) precipitates, as shown in Fig. 2d. The schematic illustration overlaid in the figure shows the relative orientation of the precipitates with respect to the Mg matrix. The $\beta_1'$ rod-like precipitates follow the $(0001)_{\beta_1'}//(11\bar{2}0)_{\alpha\text{-}Mg}$ , $[11\bar{2}0]_{\beta_1'}||[0001]_{\alpha\text{-}Mg}$ orientation relationship with the Mg lattice, as shown in Fig, 2e, and the long axis of the rod is aligned with the c-axis of the Mg lattice. The plate-shape $\beta_2'$ ($MgZn_2$) precipitates follow the $(0001)_{\beta_2'} \parallel (0001)_{\alpha\text{-}Mg}$ , $[11\bar{2}0]_{\beta_2'} \parallel [10\bar{1}0]_{\alpha\text{-}Mg}$ orientation relationship, as shown in Fig. 2f, with the plane of the plate parallel to the basal plane of the Mg lattice. The $\beta_1'$ rod-like precipitates are expected to be more effective strengthening basal slip than the plate-shape $\beta_2'$ ($MgZn_2$) precipitates. First, because the plate-shape $\beta_2'$ ($MgZn_2$) plates are distributed on the basal plane of the Mg lattice, and hence do not interact as much with the basal dislocations, as discussed before [25,26]. And second, and more importantly, because the volume fraction of the scattered plate-like precipitates is negligible in comparison with the $\beta_1'$ rod-like precipitates in the peak-aged condition selected in this work. The average length and diameter of the rod-shape precipitates were 95 ± 36 nm and 11 ± 3 nm, respectively. The volume fraction of precipitates was around 1.8 % and the average spacing between precipitates in the basal plane was 9.9 ± 4.5 nm, as estimated by the Image J software. Overall, the precipitates have similar aspect ratio (length over diameter) and dimensions than other works for Mg-2 at.% Zn alloys with similar aging conditions [10,13].

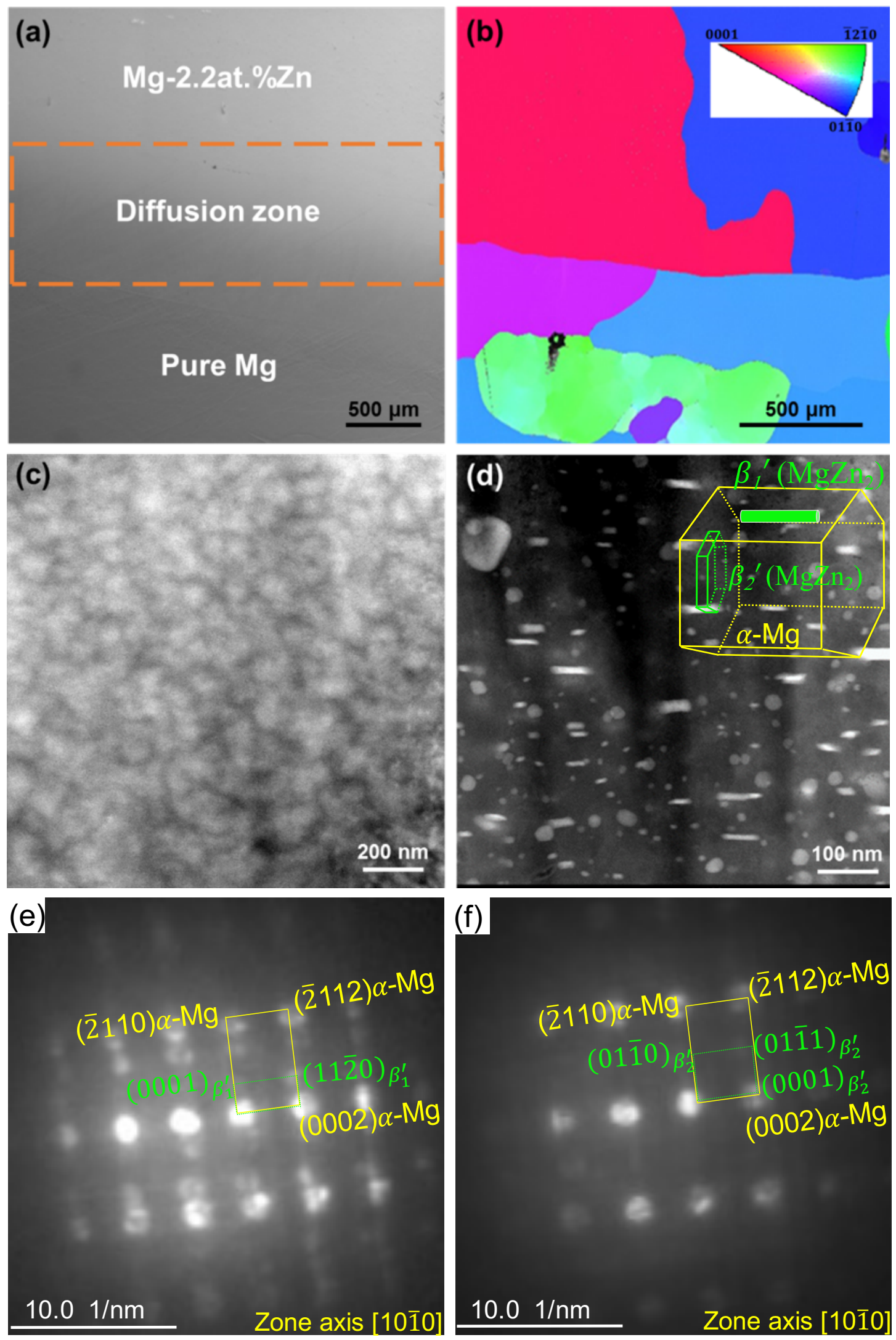

Fig. 2. (a) SEM image of the Mg /Mg-2at. %Zn diffusion couple; (b) Representative EBSD map in the diffusion couple showing the grains for micropillar compression; (c-d) HAADF STEM images of (c) Mg-2Zn AQ and (d) Mg-2Zn PA conditions, overlaid with an schematic representation of the precipitates; (e-f) Microdiffraction pattern corresponding to the (e) rod-like $\beta_1'$ precipitates and (f) plate-like $\beta_2'$ precipitates.

*3.2 Micropillar compression in grains favorably oriented for basal slip*

Micropillar compression tests were conducted at 100 °C for the grains favorably oriented for basal slip in pure Mg, Mg-2Zn-AQ and Mg-2Zn-PA alloys. The representative resolved shear stress (τ) - strain (ε) curves are plotted in Fig. 3a. The micropillars were firstly elastically deformed, up to the yield point and the CRSS determined for pure Mg was about 7 MPa, comparable to reference values [9]. The measured CRSS in Mg-2Zn-AQ was around 14 MPa, higher than that for pure Mg. In comparison, the CRSS for Mg-2Zn-PA was the highest, ≈ 52 MPa, which clearly indicates that precipitation strengthening is more effective than solid-solution strengthening for basal slip. The shape of the curves after the initial yielding was also remarkably different. In the strain hardening regime, the curves for pure Mg and Mg-2Zn-AQ were jerky, which is the common behavior observed as a result of the sudden activation of dislocation avalanches in localized slip planes. In contrast, the Mg-2Zn-PA condition showed a smooth $\tau_{RSS}$-ε curve, without sudden load drops, indicating a more "bulk-like" behavior as a result of the precipitate-dislocation interactions, because the length scale that controls the mechanical behavior in this case is the precipitate spacing, which is much smaller than the micropillar dimension. Moreover, the strain hardening rate after the initial yield was much higher than that in pure Mg and Mg-2Zn-AQ, indicating a higher rate of forest dislocation hardening [27].

The shape of the deformed pillars confirmed the activation of basal slip, as the slip traces were oriented along the basal plane in all cases, as shown in Fig. 3b. The deformed pillars of pure Mg and Mg-2Zn-AQ indicate strain localization in a discrete number of slip traces along the length of the pillar, as shown in Fig. 3b and Fig. 3c. The results suggest that the strain bursts are associated with the activation of discrete dislocation sources in adjacent basal slip planes giving rise to dislocation avalanches. They exit at the walls of the micropillars, contributing to the progressive formation of slip steps of different intensity. On the contrary, the pillars deformed in Mg-2Zn-PA in Fig. 3d display a uniform strain distribution, with only one apparent basal slip trace. This, together with the shape of the $\tau_{\mathrm{RSS}}$-$\varepsilon$ curves, suggests that the precipitates present in the Mg-2Zn-PA contribute to a higher CRSS for basal slip. More importantly, the strong interaction with the basal dislocations contributes to stop the dislocation avalanches and to diminish the localization of strain in a few slip bands. This facilitates to eliminate the jerky behavior, resulting in an increase in the storage of forest dislocations that contributes to the observed strain hardening.

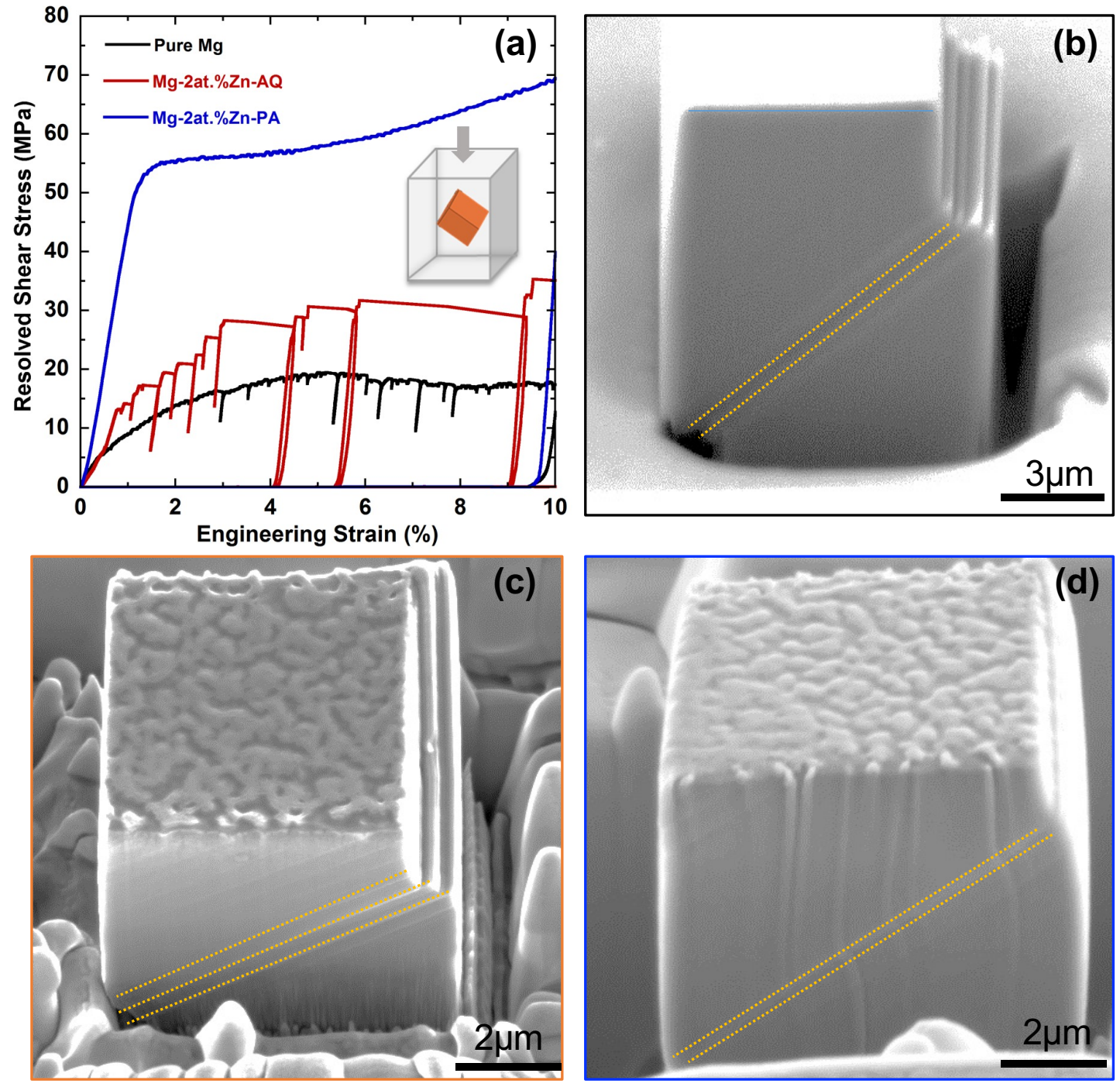

Fig. 3. (a) Representative resolved shear stress - engineering strain curves for pure Mg, Mg-2Zn-AQ and Mg-2Zn-PA along the orientation favorably oriented for basal slip. The deformed pillars corresponding to (b) pure Mg, (c) Mg-2Zn-AQ and (d) Mg-2Zn-PA, respectively.

The effect of Zn on basal slip in the Mg-2Zn-AQ condition is explored in more detail by TEM in Fig. 4. The individual basal slip traces in the deformed micropillar, observed along the $[2\bar{1}\bar{1}0]$ zone axis ($g = (0002)$), are shown in Fig. 4a. Detailed STEM images of the slip trace in Fig. 4c and Fig. 4d show that the Zn atoms tend to segregate and cluster upon interaction with the basal dislocations during deformation, which is in agreement with the solute clustering [25] and short range order [28] effects reported in Mg-Zn alloys. Solute segregation in the basal slip traces is expected to exert additional drag effects on the movement of basal dislocations. Interestingly, the BF-TEM images of Fig. 4e and Fig. 4f show the presence of an array of dislocations at different zone axis and *g* vectors, as pointed by the arrow. The imaging conditions in Fig. 4e and Fig. 4f clearly reveal that the array of dislocations correspond to <a> dislocations gliding on basal planes in response to the applied stress, since they are visible in all conditions. The analysis of the Burgers vector gives $b =< 1\bar{2}10 >$. The fact that the spacing between the dislocations in the array is uniform might indicate a low energy configuration due to their interaction with the Zn atomic clusters. It is worthy pointing out that this observation is the first experimental evidence in Mg-Zn alloys proving the drag effects of Zn segregation on the movement of basal dislocations.

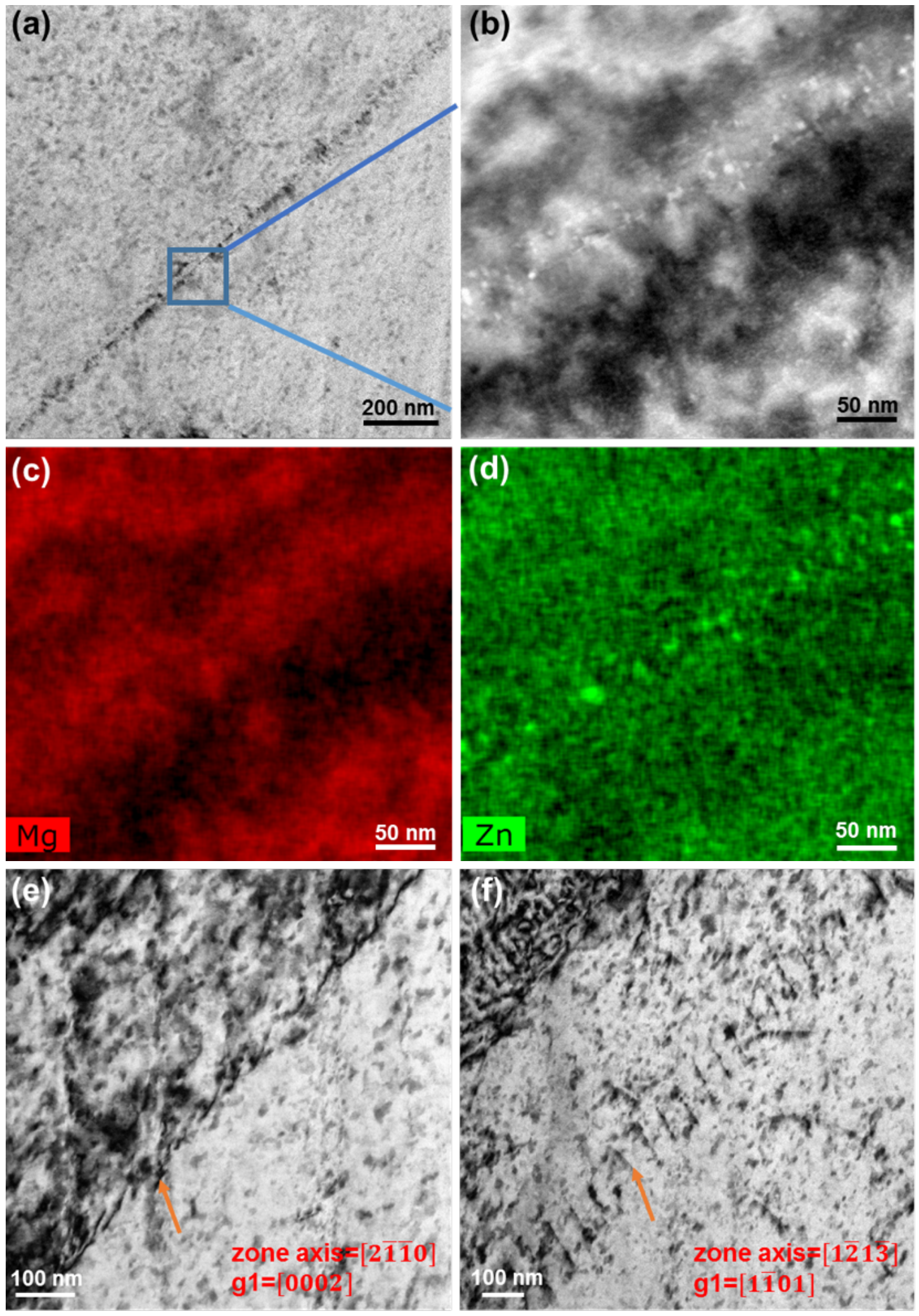

Fig. 4. TEM analysis of basal slip in the Mg-2Zn-AQ alloy: (a, b) BF-TEM images along $[2\bar{1}\bar{1}0]$ axis ($g = (0002)$) condition ; (c, d) EDS compositional maps of Mg and Zn in the selected region; (e, f) BF-TEM images of an array of dislocations observed at $[2\bar{1}\bar{1}0]$ axis ($g_1 = (0002)$) and $[1\bar{2}1\bar{3}]$ axis ($g_2 = (1\bar{1}01)$).

The precipitation effects for basal slip in the Mg-2Zn-PA condition were further studied by TEM, as shown in Fig. 5. The basal slip traces are indicated by the yellow dash lines in Fig. 5 (a-b). The rod-like $\beta_1'$ precipitates were sheared by the gliding basal dislocations. Moreover, the precipitates were transformed in globular particles around the most intense slip bands (Fig. 5b). The BF-TEM images in Fig. 5e and Fig. 5f provide further evidence for the shearing of precipitates: a high density of gliding basal dislocations is observed in the deformed micropillars, as indicated by the arrows in Fig. 5e and Fig. 5f, which in agreement with the high strain hardening rate observed in the micropillar compression tests.

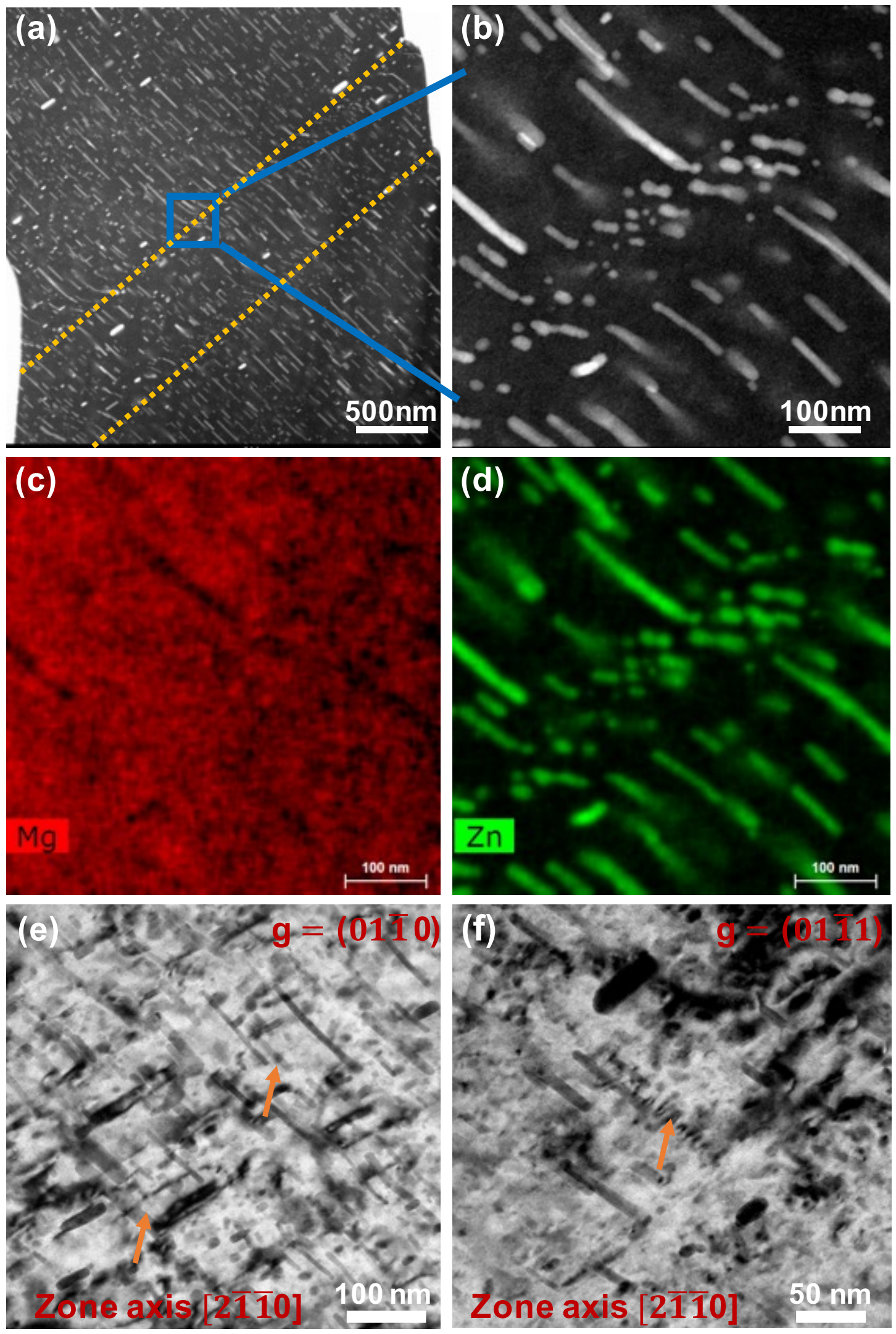

Fig. 5. TEM analysis of slip traces in Mg-2Zn-PA. (a) TEM DF image; (b) HAADF STEM image and (c-d) its corresponding EDS compositional maps: (c) Mg and (d) Zn. (e-f) TEM BF images, in the $[2\bar{1}\bar{1}0]$ zone axis, corresponding to different diffraction conditions: (e) g=$(01\bar{1}0)$ and (f) g=$(01\bar{1}1)$

Assuming that the strengthening effects are additive, the blocking effect of the rod-like $\beta_1'$ precipitates on basal slip $\tau_p$, in the Mg-2Zn-PA condition can be estimated from:

$$\tau_p = \tau_{crss} - \tau_{ss} - \tau_{Mg} \quad (1)$$

where $\tau_{crss}$ is the CRSS measured by micropillar compression, $\tau_{Mg}$ corresponds to the CRSS for pure Mg and $\tau_{ss}$ corresponds to the solid solution strengthening due to the Zn atoms that remain in solid solution after the precipitation process. The Zn elemental mapping in Fig. 5 (c-d) indicates that, after the precipitation process, the Zn content of the matrix was still around 1.7 at.%. Therefore, considering $\tau_{ss} + \tau_{Mg}$ to be similar to the CRSS measured for the Mg-2Zn-AQ condition (14 MPa), $\tau_p$ was estimated to be around 38 MPa.

A number of possible mechanisms might be responsible for the blocking effect of shearable precipitates including stacking fault energy mismatch (most relevant to the case of disordered precipitates with the same crystal structure as the matrix), interface creation, modulus mismatch, coherency strain, and anti-phase domain boundary (APB) order strengthening [29]. Due to their coherence with the matrix, order strengthening is expected to be the dominant mechanism for $\beta_1'$ ($MgZn_2$) precipitates. Anti-phase boundaries (APB) are generated if the particles are sheared by gliding dislocations and the precipitation strengthening effect can be estimated from [12,17,25–27]:

$$\tau_p = \frac{\gamma_{APB}}{2b}\left(\frac{d}{\left(\frac{0.953}{\sqrt{f}}-1\right)d} - f\right) \tag{2}$$

where the APB energy per unit area on the slip plane, $\gamma_{APB}$, represents the force per unit length opposing the dislocation cutting through the particle [29,31], $d$ is the diameter of the precipitates in the basal plane and $f$ is the volume fraction of precipitates. From the diameter of the precipitates, 11± 3 nm, and the measured volume fraction 1.8%, the APB energy $\gamma_{APB}$ was estimated to be 183 mJ/m$^2$.

### *3.3 Micropillar compression in grains favorably oriented for extension twinning*

In order to study extension twinning in Mg-Zn alloys in the solid-solution and precipitation states, micropillar compression was further conducted in pure Mg, Mg-2Zn-AQ and Mg-2Zn-PA along the $[10\bar{1}0]$ compression axis. Fig. 6a plots the resolved shear stress-strain curves. The mechanical response can be divided into three stages: first, elastic loading up to the initial yield; second, a plateau of constant stress immediately after yielding; and third, a pronounced strain hardening regime at large strains.

For pure Mg, a strain burst at peak loading was observed at a strain of ≈1%, which indicates the first twin nucleation event, as has been discussed in previous works [21,32]. The same behavior was also observed at larger strains, ≈ 2.0% in Mg-2Zn-AQ and at ≈1.2% in Mg-2Zn-PA respectively. Therefore, the CRSS for twin nucleation could be estimated from the point of the first strain burst, leading to values of 20.6 ± 0.2 MPa, 37.5 ± 1.8 MPa and 30 MPa for pure Mg, Mg-2Zn-AQ and Mg-2Zn-PA, respectively. Following twin nucleation, the stress dropped to a plateau that is controlled by the twin growth stage until the entire micropillar was twinned, as has been reported before [15,33]. The flow stress at the plateau remained relatively constant up to an applied strain of 4~6%, followed by the strain hardening regime. Therefore, the CRSS for twin growth could be estimated from the stress plateau, leading to values of 13.5 ± 0.15 MPa. 38.5 ± 2 MPa and 50 MPa for pure Mg, Mg-2Zn-AQ and Mg-2Zn-PA, respectively. It is interesting to notice that the behavior of Mg-2Zn-AQ at the stress plateau showed a much jerkier character, which might be related to the interaction of the twin boundary with the Zn solute atoms. Another worthy to mention observation is that the CRSS for twin nucleation was smaller than that for twin growth in the case of Mg-2Zn-PA, which might indicate that the precipitates offer a much higher resistance to twin growth than to twin nucleation.

Fig. 6 (b-d) shows SEM images of the deformed pillars. Since the twinning process induces the rotation of the parent orientation by approximately 90º, the orientation of the twinned regions brings the c-axis of the HCP lattice close to the loading direction. In this orientation, the twinned region is favorably oriented for pyramidal slip, but basal slip traces were also observed in pure Mg and Mg-2Zn-AQ, as indicated by the dashed lines. It is worth noticing that, in the case of Mg-2Zn-AQ, the basal slip traces in the twinned region were more diffuse than in pure Mg, as shown in Fig. 6d. Moreover, no slip traces were observed in the case of Mg-2Zn-PA. On the contrary, several twins were observed on the pillar surface in the case of Mg-2Zn-PA, as indicated by the blue arrows.

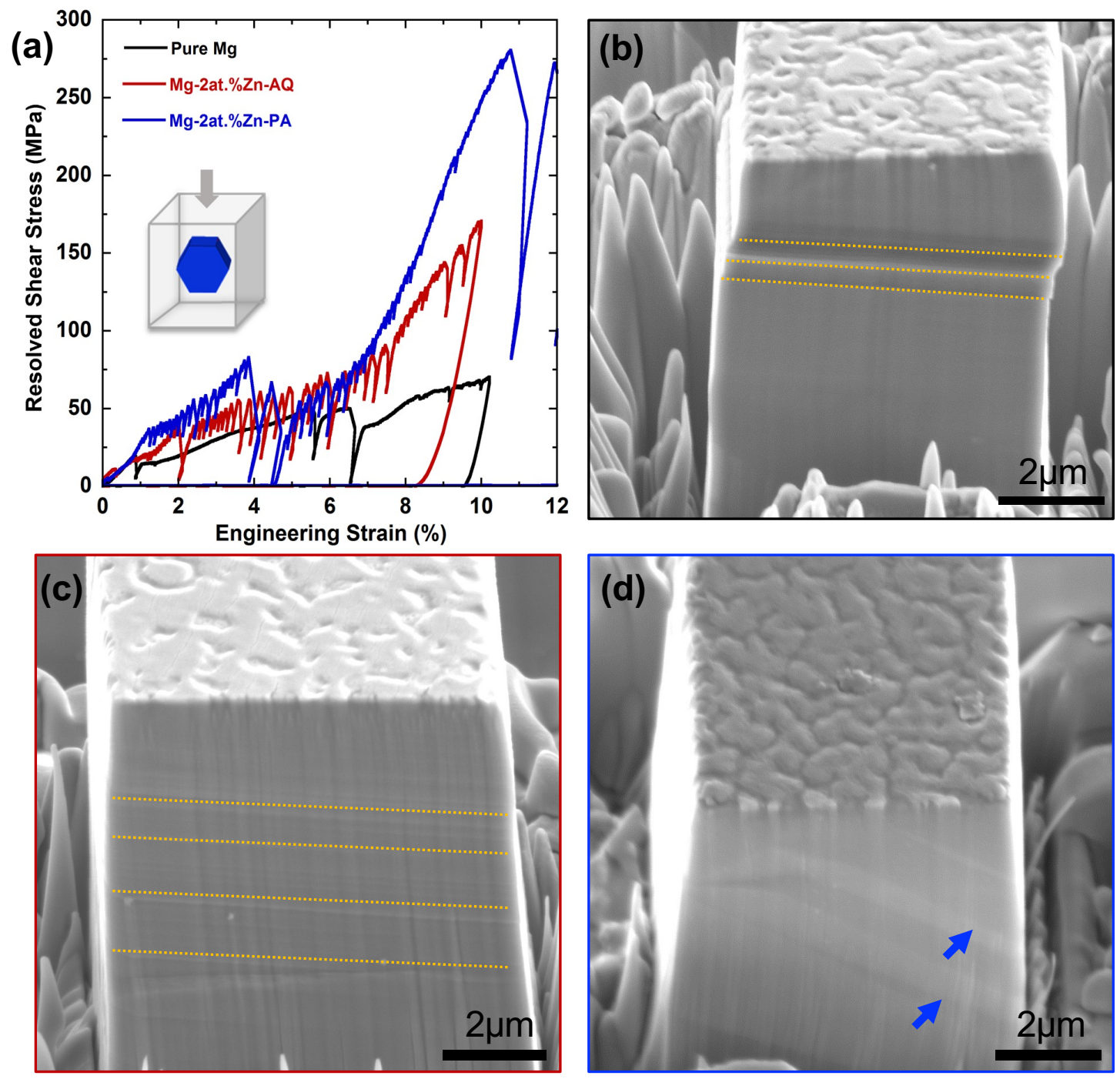

Fig. 6. (a) Representative resolved shear stress- engineering strain curves for pure Mg, Mg-2Zn-AQ and Mg-2Zn-PA.The loading axis is along the twinning favourable orientation. (b-d) Images of the deformed pillars in each condition.

To discern the twinning deformation process in each case, TEM lamellae were extracted from the deformed pillar in the Mg-2Zn-AQ condition as shown in Fig. 7. The entire pillar has re-oriented to the twin orientation (in green), as labelled in Fig. 7a, where the dashed line indicates the position of the twin boundary between the twin and the parent orientation at the bottom. Their orientation relationship analyzed by TKD is shown in Fig. 7b, confirming a rotation of 87.6° between the parent and the deformation twin, as expected for extension twinning. The basal slip traces (marked by the orange dashed lines in Fig. 7c) indicate the activation of basal slip on the twinned orientation. Further TEM analysis of the dislocation structures confirmed profuse activity of basal <a> dislocations and <c+a> pyramidal dislocations. This occurs due to the low CRSS for basal slip compared to that for pyramidal slip, despite the low SF of the former, of only 0.12 in the twinned region. Therefore, in the case of Mg-2Zn-AQ, strain is accommodated both by basal and pyramidal slip in the twinned orientation, explaining the high strain hardening regime observed after twinning.

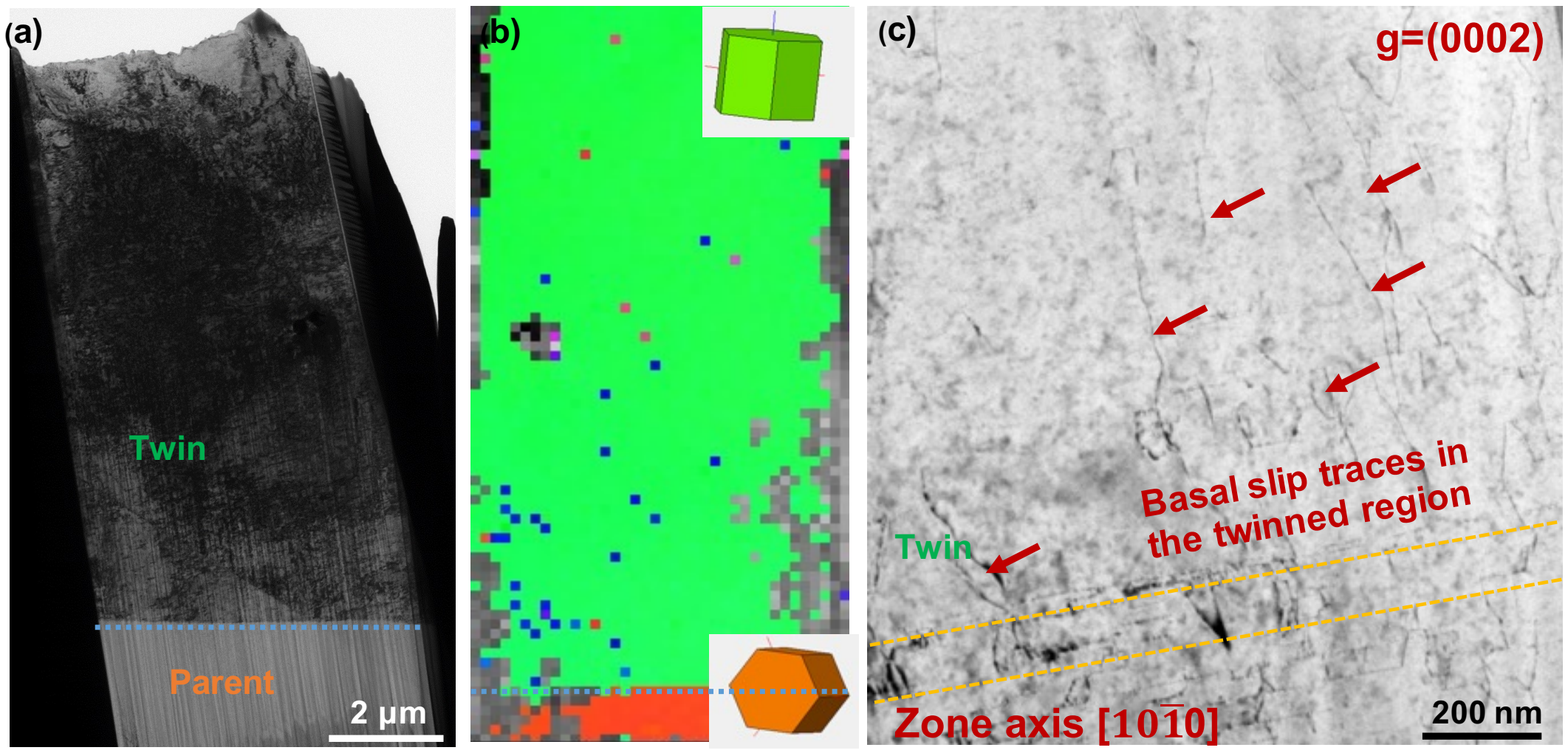

Fig. 7. (a) Longitudinal TEM image of the deformed pillar in the Mg-2Zn-AQ condition along the $[10\bar{1}0]$ direction; (b) The TKD map showing the twin and the parent orientation; (c) TEM BF image showing the dislocation structure after deformation in the diffraction condition g=(0002), showing $< c + a >$ dislocations in the twin. The basal slip traces in the twin can also be observed, as indicated by the yellow dashed line.

The effect of the precipitates on twin nucleation and growth was different for Mg-2Zn-PA, as shown in Fig. 8. Contrary to the previous cases, several twins were nucleated upon compression (Fig. 8a), of which one grew preferentially with strain, but the rest could be still observed at the bottom of the pillar, as shown in Fig. 8d and the corresponding DP in Fig. 8(e-g). This might explain why the load peak is not followed by a strain burst (Fig. 6a), as opposed to the previous cases. The misorientation between the parent and dominant twin variant (marked with a blue arrow in Fig. 8a), 86.3º, is plotted in Fig. 8b, from the TKD orientation map of Fig. 8c. It indicates that the twin has been rotated 86.3° with respect to the parent orientation.

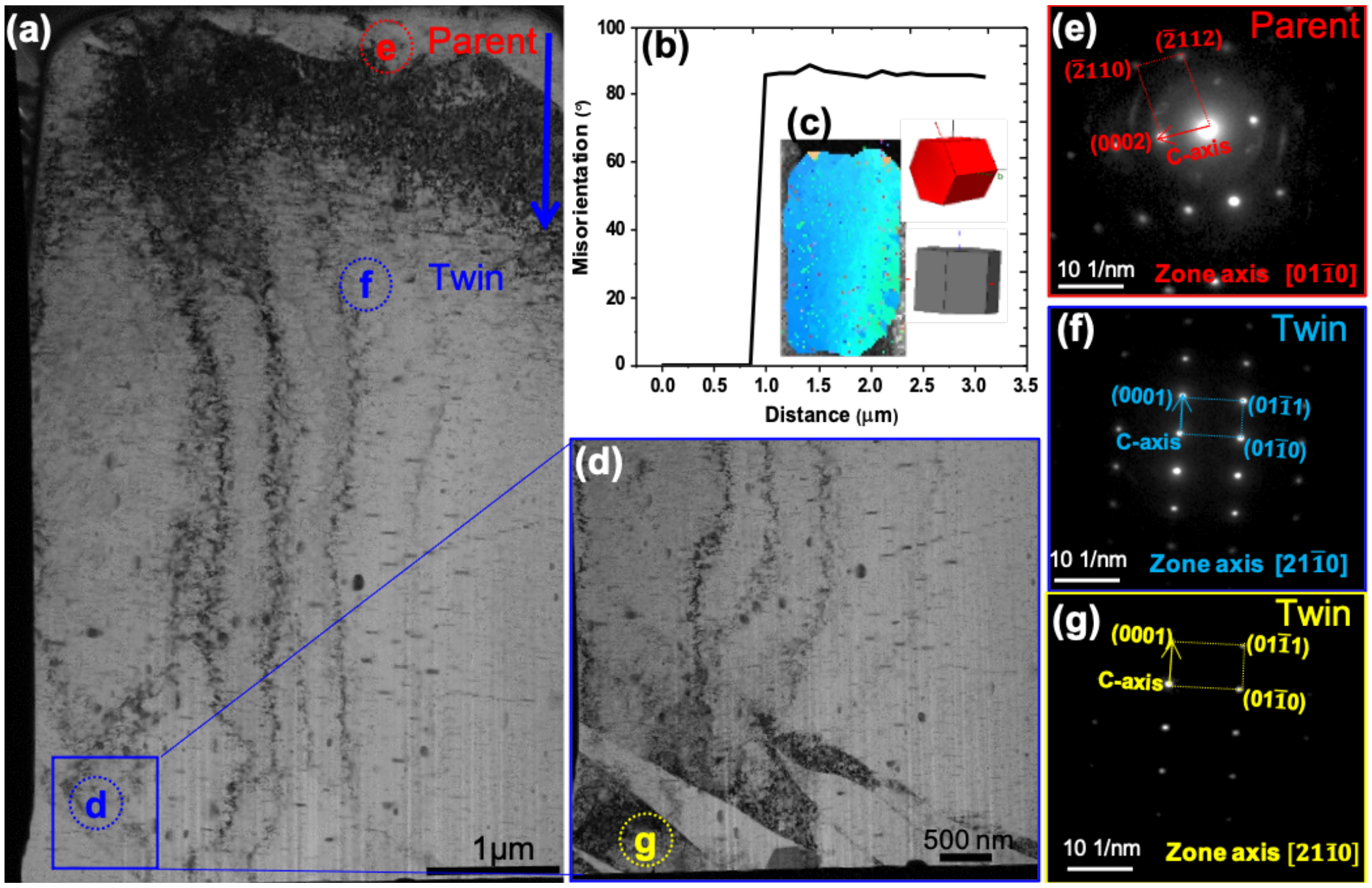

Fig. 8. (a) Longitudinal TEM image of the 7 μm deformed pillar in the Mg-2Zn-PA condition compressed along the [10-10] direction; (b) Misorientation profile across the twin boundary and (c) corresponding TKD orientation map; (d) Close-up BF-TEM image of the bottom part of the pillar where several twin variants survive; (e-g) The corresponding diffraction patterns of the parent and twins at the top and bottom part of the pillar.

The TEM images also reveal interesting observations about the interaction of the rod-shape $\beta_1'$ ($MgZn_2$) precipitates with the twin. Fig. 9a shows a close-up around the twin boundary, showing part of the parent and twin orientations, imaged along the $[10\bar{1}0]$ zone axis of the parent. In the parent orientation region, the $\beta_1'$ ($MgZn_2$) precipitates are aligned with their long axis oriented parallel to the $< c >$ axis of the Mg HCP lattice, as expected. Interestingly, the rod like $\beta_1'$ ($MgZn_2$) precipitates in the twin showed a 4º rigid body rotation with respect to those in the parent orientation. This results in the long axis of the rod precipitates being almost parallel to the basal planes in the twinned region, and therefore, the basal dislocations are not expected to shear the precipitates in the twinned region anymore. As a matter of fact, no basal slip activity was observed in the twinned regions. On the contrary, profuse pyramidal slip activity was observed in these regions, with the dislocation segments showing Orowan type bowing around the precipitates, as shown in Fig. 9b.

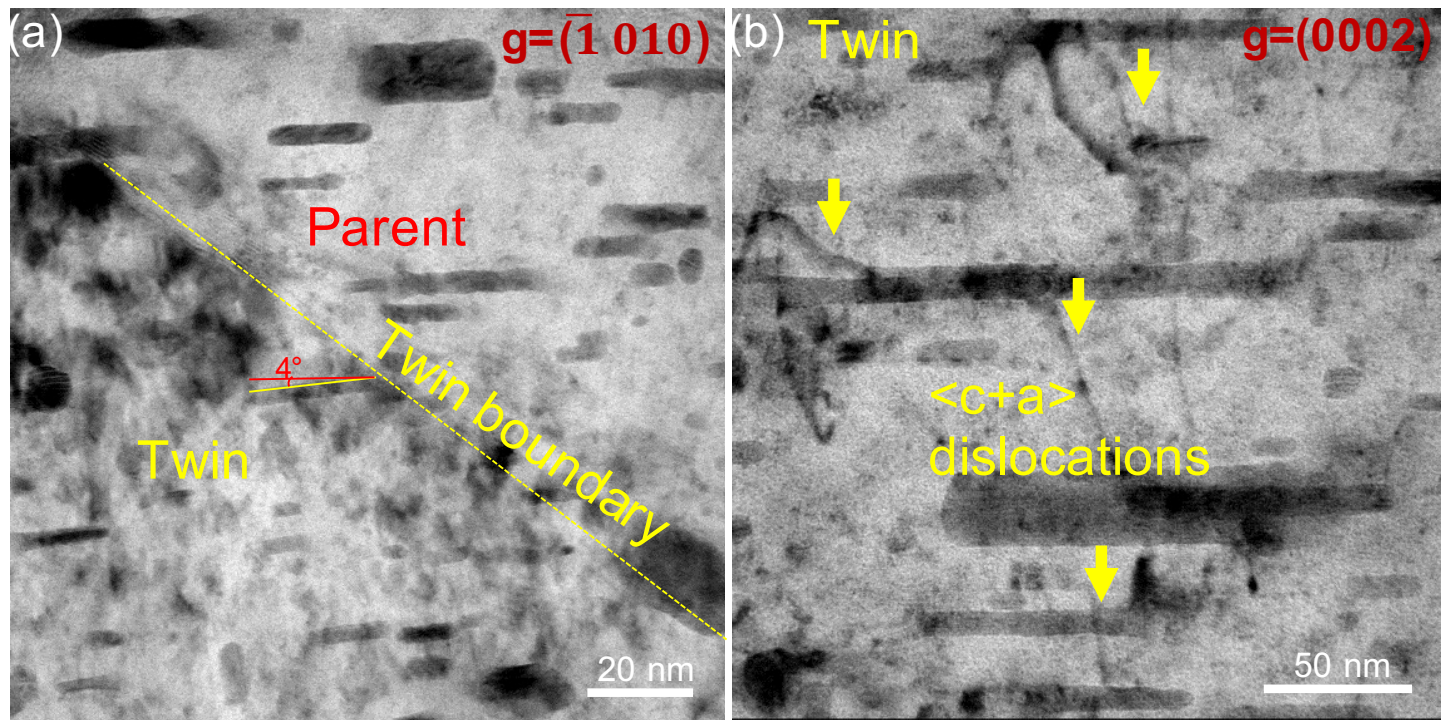

Fig. 9. (a) BF-TEM images of twin boundary region viewed along the (10$\bar{1}$0) zone axis of the parent orientation with g= ($\bar{1}$0$\bar{1}$0) condition; (b) BF-TEM image within the twinned region image along zone axis[10$\bar{1}$0] with g= (0002) condition. The yellow arrows indicate the activation of <c+a> dislocations.

*3.4 Micropillar compression in grains favorably oriented for pyramidal slip*

In the case of the grains favorably oriented for pyramidal slip (grain 3 in pure Mg and grain 6 in Mg-2at.%Zn), compressed along the <0001> direction, both pyramidal and basal slip are typically activated, as has been observed before [34,35]. This occurs because, even though the SF for pyramidal slip is significantly larger than for basal slip (Table 1), the later possesses a much lower CRSS than the former. In order to determine the CRSS for pyramidal slip unambiguously, several Mg micropillars were compressed to different levels of total strain and the sequence of activation of the different slip systems was assessed, as shown in Fig.10. The black curve shows a pillar compressed to a total strain of 10% strain, for which the activation of pyramidal <c+a> dislocations was confirmed by TEM. No large slip traces are typically formed in the case of pyramidal slip, as shown in Fig. 10b, because its activation is not followed by a large avalanche of dislocations that scape the pillar surface, as has been observed before in previous studies [34,36] The activation of pyramidal slip is followed by a large strain hardening regime, that leads to the eventual activation of basal slip, despite its low SF. This can be clearly seen in the case of the red curve, that corresponds to a pillar compressed to a total strain of 20%. At a strain of 15%, a large stress relaxation event is observed, which corresponds to the activation of basal slip and the formation of a large basal slip trace, as shown in Fig. 10c. Therefore, the initial yield point indicated by a blue star in Fig 10a can be taken as the CRSS for pyramidal slip. A similar behavior was observed in the case of Mg-2Zn in this work.

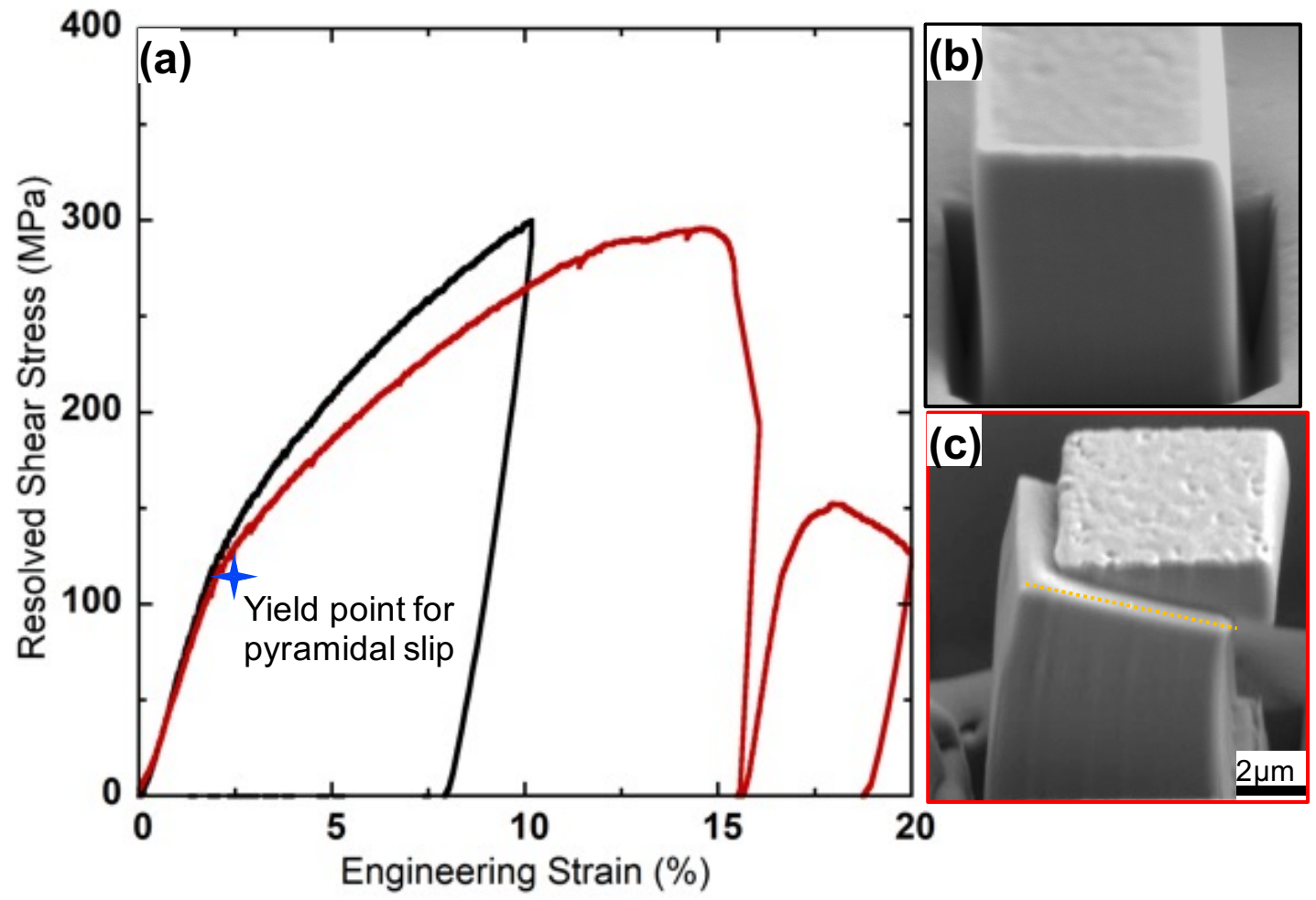

Fig. 10 (a) The resolved shear stress ~ engineering strain curves in Mg micropillars favorably oriented for pyramidal slip compressed to different levels of strain; (b) The morphology of the deformed pillar corresponding to the black curve in Fig 10a, showing no signs of basal slip activation; (c) The morphology of the deformed pillar corresponding to the red curve in Fig 10a, where a basal slip trace is clearly observed.

Fig. 11a plots the corresponding $\tau$ - $\varepsilon$ curves of Mg-2Zn-AQ,Mg-2Zn-PA and pure Mg. Based on the above, pure Mg shows a CRSS for pyramidal slip of 120 ± 5 MPa, followed by a strong strain hardening regime. The CRSS for pyramidal slip in the case of Mg-2Zn-AQ raised to 200 MPa, while in the case of Mg-2Zn-PA was of the order of 240 MPa, suggesting that Zn alloying offers a similar solid solution and precipitation strengthening effect in the case of pyramidal slip. Interestingly, the deformed pillar in Mg-2Zn-AQ in Fig. 11c shows both pyramidal (red lines) and basal (yellow lines) slip traces, while the deformed pillar in Mg-2Zn-PA shown in Fig. 11d, only shows pyramidal slip traces (in red). The fact that no basal slip activity was found in the latter case agrees with the strong precipitation hardening observed for basal slip in this case, which prevents its activation.

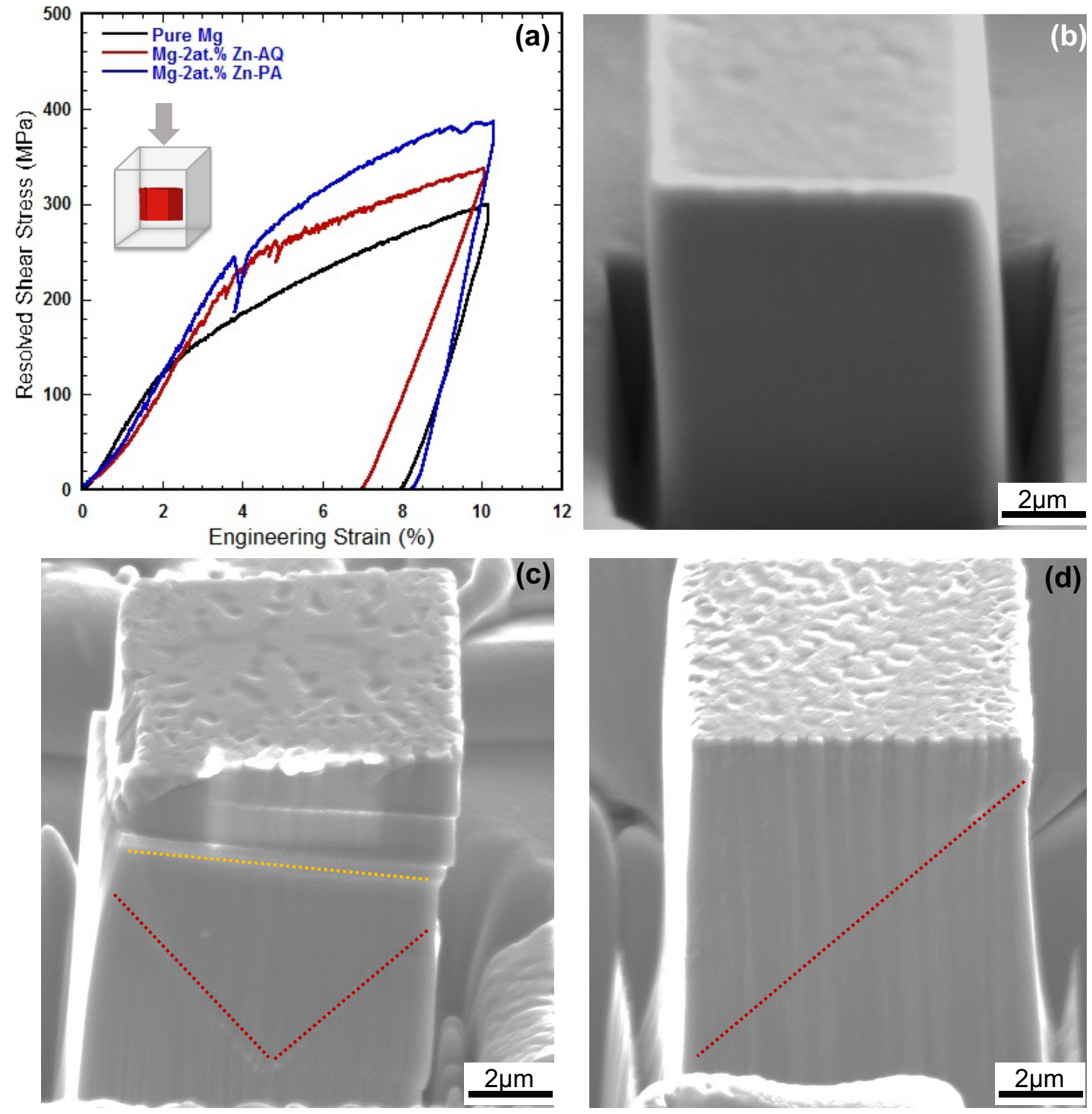

Fig. 11. (a) Representative resolved shear stress-engineering strain curves for pure Mg, Mg-2Zn AQ and Mg-2Zn PA favorably oriented for pyramidal slip. SEM images of the deformed pillars in (b) Mg, (c) Mg-2Zn AQ and (d) Mg-2Zn PA. The scale bar is 2 μm in all cases.

### *3.5 Solid solution and precipitation effects in the anisotropy of Mg-Zn alloys*

Finally, Fig. 12a summarizes the CRSS for each individual deformation mode (basal slip, twin nucleation, twin growth and pyramidal slip) for pure Mg and solid-solution and precipitation strengthened Mg-Zn alloys. It is clear that the $\beta_1'$ ($MgZn_2$) precipitates exert a higher strengthening than Zn in solid solution, especially with respect to the soft slip systems, i.e., basal slip and twin growth. This has important consequences on the plastic anisotropy in each case. To assess this, Fig. 12b plots the CRSS ratios of pyramidal slip, twin nucleation and twin growth with respect to the softest slip system, basal slip. Results from literature for Mg-Zn [37] and other alloys, like Mg-Al [33,38], are also included for comparison. The blue and green shadowed areas act as a guide to the eye, representing the precipitation and solid solution strengthened alloys, respectively. It is clear that, even though solid solution strengthening contributes to the hardening of all deformation modes, it does not contribute to reduce the anisotropy ratios noticeably with respect to pure Mg, because the pyramidal to basal CRSS ratios remain always above 10. However, the CRSS ratios are reduced below 5 for the precipitation strengthened alloys, mostly because the hardening of basal slip is considerably higher than that found for pyramidal slip. Similar results with respect to a more efficient reduction of the anisotropy ratios in precipitation strengthened alloys with respect to solid solution strengthened ones have also been found in the case of Mg-Al, as shown in Fig. 12b.

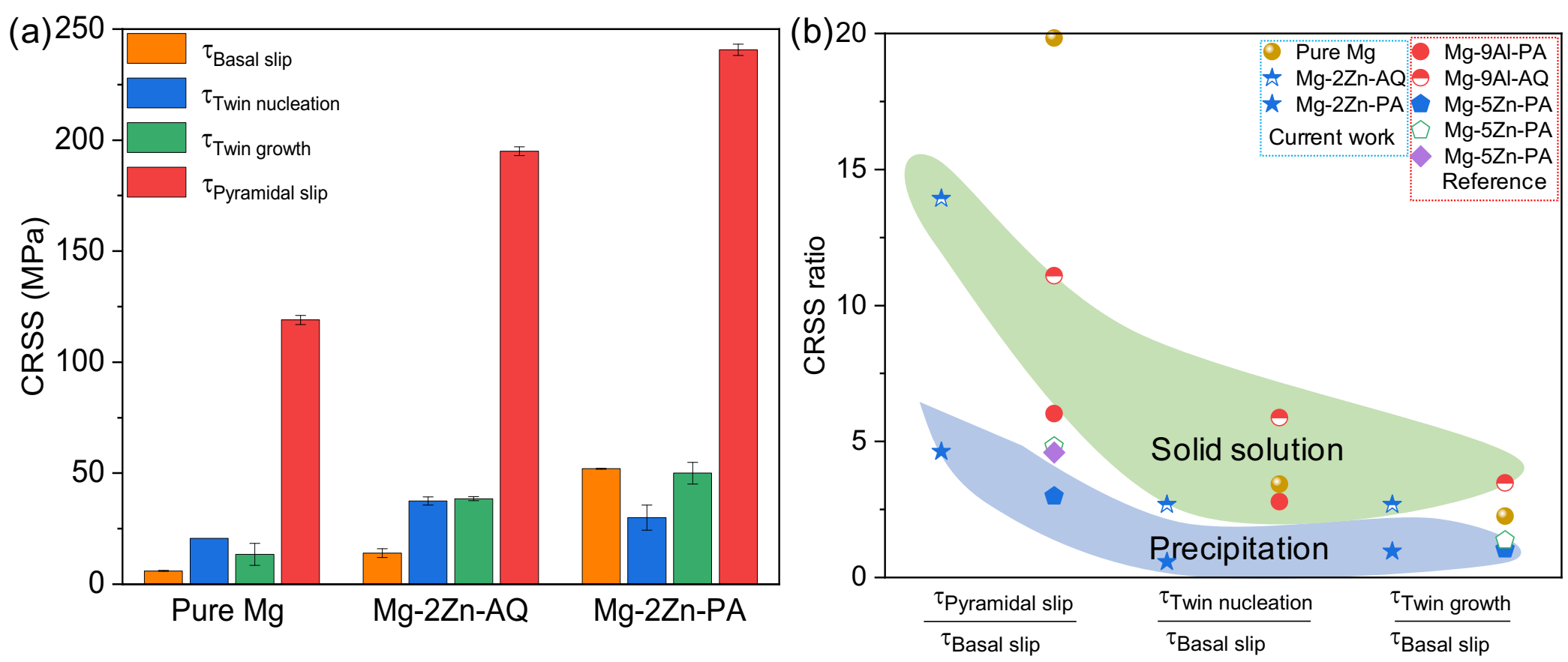

Fig. 12. (a) The resolved shear stress basal slip, twin nucleation, twin growth, and pyramidal slip for pure Mg, Mg-2Zn-AQ and Mg-2Zn-PA respectively. (b) The CRSS ratios of the non-basal slip mode (pyramidal slip, twin nucleation, twin growth) to the basal slip. Reference data are also included [13,33,37–39].

## 4. Conclusions

Solid solution and precipitation strengthening effects on the CRSS for basal slip, extension twinning, and pyramidal slip have been evaluated in Mg-2at.% Zn alloys in the as-quenched (AQ) and peaked-aged (PA) conditions using diffusion couples and micropillar compression. The key findings can be summarized as follows:

- The Zn solute atoms are homogenously dispersed in solid solution in the AQ state. Peak aging at 200ºC for 10 h results in the formation of rod-shape precipitates of $\beta_1'$ ($MgZn_2$) and plate-shape $\beta_2'$ ($MgZn_2$). The fraction volume of the scattered plate-like precipitates is negligible in comparison with $\beta_1'$ rod-like precipitates . The rod-shape precipitates have a mean diameter and length of 11±3 nm and 95 ± 36 nm, respectively, and with their long axis aligned along the <c> axis of the α-Mg matrix.
- Zn clustering seems to play an important role on the solid solution strengthening of basal slip for supersaturated Mg-Zn alloys, providing a strengthening of around 7 MPa for Mg-2 at.% Zn, with respect to pure Mg. Solid solution is more effective strengthening the CRSS for twin growth than for basal slip, presumably due to Zn segregation at the twin boundaries. The strengthening effect is of the order of ≈ 25 MPa with respect to pure Mg, for a Zn content of 2 at.%.

- The precipitation of rod-like $\beta_1'$ ($MgZn_2$) precipitates, with a diameter of 11±3 nm and a volume fraction of 1.8%, is very effective strengthening basal slip in Mg-Zn alloys, providing an increase of 36 MPa in the CRSS for basal slip with respect to Mg-2 at.% Zn in solid solution. The $\beta_1'$ ($MgZn_2$) precipitates are sheared by the basal dislocations and order strengthening is expected to be the dominant mechanism responsible for the precipitation hardening. The $\beta_1'$ ($MgZn_2$) precipitates favor twin nucleation with respect to the solid-solution condition, but they induce a strong pinning effect on twin boundary migration, presumably because the precipitates loss the coherency with the HCP lattice upon twinning as they do not rotate with the lattice. The strengthening effect of $\beta_1'$ ($MgZn_2$) precipitates on twin growth was estimated to be around 12 MPa with respect to the CRSS for twin growth for Mg-2 at.% Zn in solid solution.
- Solid solution and precipitation strengthening in pyramidal slip increase the CRSS to 200 MPa and 240 MPa, respectively, in Mg-2 at.% Zn with respect to the 120 MPa found for pure Mg. As a result, the anisotropy ratios remain high and similar to pure Mg in the solid solution strengthened Mg-Zn alloys. However, they are substantially reduced in precipitation strengthened Mg-Zn alloys.

## Acknowledgements

The research leading to these results has received funding from the Madrid Region under programme (S2018/NMT-4381), MAT4.0 project (Spain) and from the Spanish Ministry of Science and Innovation (PID2019-109962RB-I00). The EPMA measurements were carried out in the Centro Nacional de Microscopía Electrónica of the Universidad Complutense de Madrid. The authors Na Li and Chuanyun Wang are grateful to the China Scholarship Council (grant number: 201506020081 and 201406290011) for financial support.